\documentclass[3p,times]{elsarticle}
\usepackage{graphicx,booktabs,amsmath}
\journal{arXiv}

\begin{document}
\begin{frontmatter}
\title{A trust-based recommendation method using network diffusion processes}

\author[inst1,inst2]{Ling-Jiao Chen}
\ead{lingjiao.chen@foxmail.com}
\author[inst1,inst2]{Jian Gao}
\ead{gaojian08@hotmail.com}
\address[inst1]{CompleX Lab, Web Sciences Center, University of Electronic Science and Technology of China, Chengdu 611731, China}
\address[inst2]{Big Data Research Center, University of Electronic Science and Technology of China, Chengdu 611731, China}

\begin{abstract}
A variety of rating-based recommendation methods have been extensively studied including the well-known collaborative filtering approaches and some network diffusion-based methods, however, social trust relations are not sufficiently considered when making recommendations. In this paper, we contribute to the literature by proposing a trust-based recommendation method, named CosRA+T, after integrating the information of trust relations into the resource-redistribution process. Specifically, a tunable parameter is used to scale the resources received by trusted users before the redistribution back to the objects. Interestingly, we find an optimal scaling parameter for the proposed CosRA+T method to achieve its best recommendation accuracy, and the optimal value seems to be universal under several evaluation metrics across different datasets. Moreover, results of extensive experiments on the two real-world rating datasets with trust relations, Epinions and FriendFeed, suggest that CosRA+T has a remarkable improvement in overall accuracy, diversity and novelty. Our work moves a step towards designing better recommendation algorithms by employing multiple resources of social network information.
\end{abstract}

\begin{keyword}
Recommender system \sep Trust relations \sep Vertex similarity \sep Network diffusion \sep Complex networks
\end{keyword}

\end{frontmatter}

\section{Introduction}
The fast development of information technologies has spawned the emergence of the E-commerce and largely boosted its expansion during the past decades \cite{Schafer1999,Lu2012}, especially in China along with its rapid economic growth \cite{Clemes2014,Gao2018}. Recently, a large variety of online stories and services (e.g., online books, music, movies, etc) have made our lives much easier, however, the tremendous amount of available information in the era of big data has caused a serious problem of information overload \cite{Chen2009}. For example, it will be extremely hard for us to review thousands of online stores before choosing a box of favourite chocolate. To address this problem, recommender systems as an information filtering technology have been widely applied by online platforms \cite{Liu2009b,Bobadilla2013}, where users are provided customized and personalized services. Back to the example, recommender systems will automatically recommend us a few boxes of chocolate that may meet our tastes after analyzing our historical behavior big data including purchase records and search archive \cite{Zhang2012,Jin2013} that are recorded by various socioeconomic platforms \cite{Chen2014,Gao2016}.

As the core of recommender systems, a variety of recommendation algorithms have been proposed and applied to online platforms. One of the most well-known methods is the collaborative filtering (CF) \cite{Resnick1994} including the user-based CF (UCF) and the item-based CF (ICF), in which items of potential interest are recommended based on the similar users' opinions and the similarity between items, respectively \cite{Sarwar2001}. Later, some dynamical processes borrowed from the field of statistical physics are introduced into the design of diffusion-based recommendation algorithms including the heat conduction (HC) method \cite{Zhang2007}, the mass diffusion (MD) method \cite{Zhang2007b}, a hybrid method of HC and MD \cite{Zhou2010}, a weighted HC method \cite{Liu2011}, and some others \cite{Lu2011,Liu2012,Zeng2014}. The network-based diffusion is indeed a resource-allocation (RA) process \cite{Ou2007}, for example, MD is essentially a two-step RA process on ``user-object'' networks \cite{Zhou2007}. Within this framework, recent literature proposed a method built using the CosRA index \cite{Chen2017}, which combines advantages of both the cosine similarity and the RA index. In CosRA, resources are initialized for each object and then redistributed via the CosRA-based transformation. Other network-based methods are reviewed by the recent survey paper \cite{Yu2016}.

Most of these aforementioned recommendation algorithms are designed solely based on the users' rating information \cite{Lim2010}, however, the relationships among users (e.g., trust relations \cite{Massa2007,Ma2009}) embedded in social networks are always ignored \cite{Ellison2007,Yuan2016}. Yet, in real-world observations, our preference of products or adoption of information could be also affected by our social relationships \cite{Centola2010,Gao2015b} such as the friends in working places and the people connected through social media or by mobile phones \cite{Wang2016}. Intuitively, we are more likely to adopt a trusted friend's suggestions than those coming from a stronger in online communities \cite{Liao2012}. To this point, some recent works have integrated trust relations into recommender systems \cite{ODonovan2005,Walter2008}. For example, Jamali et al. \cite{Jamali2009} proposed a random walk method that combines trust-based \cite{Andersen2008} and item-based CF approaches, Ma et al. \cite{Ma2011} proposed a social trust ensemble framework that fuses both users' tastes and trusted friends' favors, Shen et al. \cite{Shen2015} utilized a trust-combined user-based CF approach by proposing two user trust models, and Guo et al. \cite{Guo2016} proposed a trust-based matrix factorization technique that integrates ratings and multiple trust information. However, studies on integrating trust relations into the diffusion-based methods remain still insufficiently \cite{Wang2017}, which urges further explorations on designing better methods under the network-based diffusion framework by leveraging multiple resources of social information.

In this paper, we propose a trust-based recommendation method, named CosRA+T, by introducing the trust relations among users into the resource-allocation processes of the original CosRA method. Specifically, the amount of the resources received by trusted users are scaled by a tunable parameter before the following redistribution back to their collected objects. Interestingly, we find an optimal value of the scaling parameter for the best recommendation accuracy under several evaluation metrics across different datasets, suggesting the universality of the optimal scaling in implicating the proposed CosRA+T method. Further, we employ two real-world rating and trust datasets, Epinions and FriendFeed, to comprehensively test the performance of CosRA+T and compare it with five benchmark methods. Results suggest that CosRA+T improves the overall performance by giving a highly accurate, inspiring diverse and well novel list of recommendations. Our work highlights the role that social trust relations play in enhancing the algorithmic performance of trust-based recommendation methods.

The remainder of this paper is organized as follows. Section~\ref{sec2} introduces some benchmark recommendation methods and the proposed CosRA+T method. Section~\ref{sec3} introduces the used datasets and performance evaluation metrics. Section~\ref{sec4} presents our main results. Finally, Section~\ref{sec5} provides conclusion remarks and related discussions.

%%%%%%%%%%%%%%%%%%%%%%%%%%%%%%%%%%%%%%%%%%
\section{Methods}
\label{sec2}

In this section, we will first introduce some basic notations that are traditionally used in describing online rating systems and recommendation algorithms. Then, we will briefly describe the five benchmark methods. Last, we will introduce the proposed CosRA+T method in detail.

\subsection{Notations and Benchmark Methods}
Online rating systems can be modeled by a ``user-object'' bipartite network $G(U, O, E_R)$, where $U=\{u_{1}, u_{2}, \ldots, u_{m}\}$ is the set of users, $O=\{o_{1}, o_{2}, \ldots, o_{n}\}$ is the set of objects, and $E_R =\{e_1, e_2, \ldots, e_l\}$ is the set of rating links. The bipartite rating network can be naturally represented by a adjacency matrix $A$, where the element $a_{i\alpha}=1$ if there is a link connecting user $i$ and object $\alpha$, and the element $a_{i\alpha}=0$ if otherwise. The Greek and Latin letters are used to distinguish object-related and user-related indices, respectively. The key purpose of recommendation algorithms is to provide a list of $L$ objects in the recommendation list $o_i^L$ for the target user $i$.

Five benchmark recommendation methods will be introduced in the following, including global ranking (GR), user-based collaborative filtering (UCF), heat conduction (HC), mass diffusion (MD), and CosRA. Based on the observations that users prefer popular objects, GR \cite{Zhou2007} as the most straightforward method recommends objects with the largest degree after sorting all objects in the descending order according to their degrees. In UCF, users will be recommended the objects collected by the users who share the similar tastes, where the user similarity is usually measured by the cosine similarity \cite{Liu2009}. Together, there is also the item-based collaborative filtering (ICF), where similar objects to the users' past collected ones will be recommended.

Both MD and HC can be considered as network-based resource-allocation processes. For a given user $i$, the initial resource of all objects is denoted by the vector $f^{(i)}$, where $f_{\alpha}^{(i)}=1$ if user $i$ has collected object $\alpha$, and $f_{\alpha}^{(i)}=0$ if otherwise. Then, these resources are reallocated via the transformation ${f'}^{(i)}=Wf^{(i)}$, where $W$ is the resource transfer matrix and ${f'}^{(i)}$ is the vector of the final resource. The resource transfer matrices $W$ in MD and HC are mutually transposed \cite{Zhou2010}. Specifically, the element of $W$ in MD is given by \cite{Zhou2007}
\begin{equation}
    w_{\alpha \beta} = \frac{1}{k_\beta} \sum_{i=1}^{m} \frac{a_{i\alpha}a_{i\beta}}{k_i},
\end{equation}
and the element of $W$ in HC is given by \cite{Zhou2010}
\begin{equation}
    w_{\alpha \beta} = \frac{1}{k_\alpha} \sum_{i=1}^{m} \frac{a_{i\alpha}a_{i\beta}}{k_i},
\end{equation}
where $k_\alpha$ and $k_\beta$ are respectively the degrees of objects $\alpha$ and $\beta$, $k_i$ is the degree of user $i$, and $m$ is the total number of users. So far, there have been many variants of the original resource transfer matrices and their associated recommendation algorithms \cite{Yu2016}.

CosRA is a recently proposed recommendation method based on the CosRA similarity index, which combines both the cosine index and the resource-allocation (RA) index \cite{Ou2007}. Specifically, for a pair of objects $\alpha$ and $\beta$, the CosRA index is given by \cite{Chen2017}
\begin{equation}
S_{\alpha\beta}^{CosRA} = \frac{1}{\sqrt{k_{\alpha} k_{\beta}}} \sum_{i=1}^{m} \frac{a_{i\alpha} a_{i\beta}}{k_i}.
\end{equation}
In CosRA, the resource of object $\alpha$ is initialized as $f_{\alpha}^{(i)}= a_{i\alpha}$ for a given user $i$. Then, these resources are redistributed via the transformation ${f'}^{(i)}= S^{CosRA} f^{(i)}$, where ${f'}^{(i)}$ is the $n$-dimensional vector recording all the final resources that located on each object. After sorting all objects by their final resources $f'^{(i)}$, the top-$L$ uncollected objects are recommended to the user $i$.

\subsection{Trust-Based Recommendation Method}

The trusted-based CosRA+T method is proposed by introducing the trust relations into the network-based diffusion processes of the original CosRA method. The intuition behind the new method is that not only the similarity among objects or users can help predict users' preferences to objects, but also the friendship or trust relations could potentially affect users' decisions on choosing objects. For example, if two users have the similar tastes as indicated by their past ratings, the performance of recommendations to either of them may be further improved if they are close friends who trust each other and have similar demands in the same living environment.

\begin{figure}[t]
  \centering
  \includegraphics[width=0.9\textwidth]{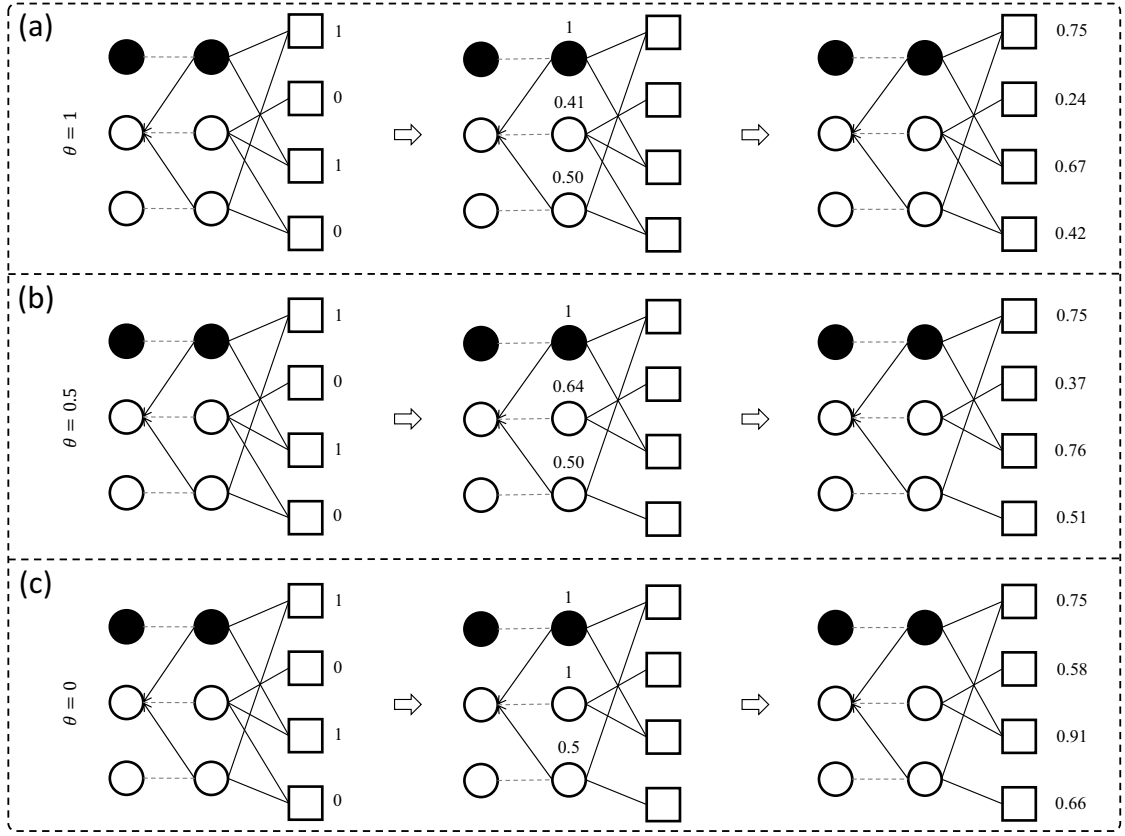}
  \caption{Illustrations of the CosRA+T method. Initially, for a target user (colored black), the resource of each object is assigned by Eq.~(\ref{eq:int}). Then, users receive resources from the objects that they have collected by Eq.~(\ref{eq:user}). Finally, users redistribute resources to their collected objects after considering the trust relations regarding the target user. Only for users (the second user in this example) trusted by the target user (the first user), their resources are scaled by a tunable parameter $\theta$ before redistributed by Eq.~(\ref{eq:object}). Values besides the nodes are the amount of received resources. Panel (a) illustrates the case of $\theta = 1$, where the CosRA+T method degenerates into the original CosRA method. Panels (b) and (c) illustrate the case of $\theta = 0.5$ and the case of $\theta = 0$, respectively.}
  \label{Fig1}
\end{figure}

To explore the role that trust relations play in enhancing or suppressing the recommendation performance, we propose the CosRA+T method by taking the ``user-user'' trust network into consideration. The trust from user $i$ to $j$ defines a link from node $i$ to $j$. Formally, the trust network can be represented by a adjacency matrix $B$, where the element $b_{ij}=1$ if there is a link from node $i$ to $j$, and $b_{ij}=0$ if otherwise. Together, Figure~\ref{Fig1} illustrates the ``user-object'' bipartite network and the ``user-user'' trust network, where circles and squares correspond to users and objects, respectively. Note that, two circles in the same row connected by the dashed lines correspond to the same user. The trust relations are presented by solid directed and unweighted links from users (circles) in the right column to their trusted users (circles) in the left column.

The CosRA+T method works in three steps: First, for a given user $i$, the resource that object $\alpha$ is initially assigned is given by
\begin{equation}
f_{\alpha}^{(i)}= a_{i\alpha},
\label{eq:int}
\end{equation}
where $a_{i\alpha}=1$ if user $i$ has collected object $\alpha$, and $a_{i\alpha}=0$ if otherwise.
Second, user $i$'s neighboring users receive resources diffused through the bipartite network from their collected objects. Formally, the resource that user $j$ receives is given by
\begin{equation}
f_{j}^{(i)}= \sum_{\alpha=1}^{n} \frac{a_{j\alpha}}{\sqrt{k_{j}k_{\alpha}}} f_{\alpha}^{(i)},
\label{eq:user}
\end{equation}
where $k_\alpha$ and $k_j$ are respectively the degree of object $\alpha$ and user $j$, and $n$ is the total number of objects. Third, users redistribute their resources back to all objects after considering the trust relations from the target user $i$. Specifically, the amount of user $j$'s resources are scaled by a tunable parameter $\theta$ before the redistribution if user $j$ is trusted by the user $i$, otherwise the resources of user $j$ are directly redistributed back to the collected objects. Formally, the final resource that object $\beta$ receives regarding the target user $i$ is given by
\begin{equation}
{f'}_{\beta}^{(i)}= \sum_{j=1}^{m} \frac{a_{j\beta}}{\sqrt{k_{j}k_{\beta}}} (b_{ij} f_{j}^{\theta(i)} + (1-b_{ij}) f_{j}^{(i)}),
\label{eq:object}
\end{equation}
where $b_{ij}=1$ if user $i$ trusts user $j$ and $b_{ij}=0$ if otherwise, $\theta$ is a tunable scaling parameter, and $m$ is the total number of users. After sorting all objects by their final resources ${f'}^{(i)}$, the top-$L$ uncollected objects are the recommendations to the target user $i$.

As the value of the scaling parameter $\theta$ increases from 0 to 1, the effects of the trust information in CosRA+T gradually diminish. In the case of $\theta=0$ as illustrated in Figure~\ref{Fig1}(a), the resources of the trusted users become 1 no matter how many resources they receive while the resources of the untrusted users remain unchanged before the redistribution. In the case of $\theta=0.5$ as illustrated in Figure~\ref{Fig1}(b), the resources of the trusted users are squared and thus increased before the redistribution. In the case of $\theta=0$ as illustrated in Figure~\ref{Fig1}(c), the trust relations have no effects on the resource-allocation processes, and CosRA+T degenerates into the original CosRA method.

%%%%%%%%%%%%%%%%%%%%%%%%%%%%%%%%%%%%%%%%%%
\section{Data and Metrics}
\label{sec3}

In this section, we first introduce the two rating datasets with the information of the trust relations among users, based on which we implement our CosRA+T method. Then, we introduce the evaluation metrics for testing the performance of recommendation methods.

\subsection{Rating and Trust Datasets}

Two benchmark rating datasets, namely, Epinions and FriendFeed, are used to test the recommendation performance. Epinions is a general consumer review website where people can review products by writing subjective reviews while FriendFeed is a social networking and bookmarking website where people can rate and share customized feeds. Both datasets use a 5-point rating scale from 1 to 5 (i.e., worst to best). For building the ``user-object'' rating network, ratings are converted to binary links by assigning 1 if the rating value is no less 3 and 0 if otherwise. After the coarse-graining, the Epinions dataset contains 4066 users, 7649 objects and 154122 rating links with the network sparsity $S_R = 4.96\times 10^{-3}$, and the FriendFeed dataset contains 4148 users, 5700 objects and 96942 rating links with the network sparsity $S_R = 4.10\times 10^{-3}$.

\begin{table}[t]
  \centering
  \footnotesize
  \caption{Summary statistics of the two rating and trust datasets.}
    \begin{tabular*}{\textwidth}{@{\extracolsep{\fill}}lcccccc}
    \toprule
    Dataset & Users ($m$)  & Objects ($n$) & Rating Links ($l_R$) & Sparsity ($S_R$)    & Trust Links ($l_T$) & Sparsity ($S_T$) \\
    \midrule
    Epinions & 4,066  & 7,649  & 154,122 & 4.96$\times 10^{-3}$ & 217,071 & 1.31$\times 10^{-2}$ \\
    FriendFeed & 4,148  & 5,700  & 96,942 & 4.10$\times 10^{-3}$ & 386,804 & 2.25$\times 10^{-2}$ \\
    \bottomrule
    \end{tabular*}
    \begin{flushleft}
    \emph{Notes}: The $S_R= l_R/(m \times n)$ is the rating network sparsity, and the $S_T=l_T/(m\times m)$ is the trust network sparsity.
    \end{flushleft}
  \label{Tab1}
\end{table}

The two datasets contains also information of social networks, which can be used to estimate trust relations among users. On the two platforms (Epinions and FriendFeed), users can be friends by following each other in addition to rating objects. For building the ``user-user'' trust network, a directed binary link from node $i$ to $j$ is assigned 1 if user $i$ follows user $j$ and 0 if otherwise. Epinions dataset contains 217,071 trust links with the network sparsity $S_T = 1.31\times 10^{-2}$, and FriendFeed dataset contains 386,804 trust links with the network sparsity $S_T = 2.25\times 10^{-2}$. Basic statistics of the Epinions and FriendFeed datasets are summarized in Table~\ref{Tab1}.

\subsection{Evaluation Metrics}

We employ a 10-folder cross-validation strategy to evaluate the algorithmic performance in each independent realization. Specifically, all ratings are divided into 10 equal sized subsamples. Then, one subsample is left out as the testing set and the remaining 9 subsamples are used as the training set. Next, the cross-validation process is repeated 10 times, making sure that each subsample serves as the testing set once. Finally, the 10 results are averaged to produce one single result for this independent realization. To quantitatively compare the recommendation performance, we apply eight commonly used evaluation metrics, including five accuracy metrics (AUC, Ranking Score, Precision, Recall, and $F_1$), two diversity metrics (Hamming distance and Intra-similarity), and one novelty metric (Popularity). These metrics are respectively introduced in the following.

Accuracy is the most important metric for the recommendation performance evaluation. We start by introducing two accuracy metrics that are independent of the recommendation list's length $L$. One is the $AUC$ (area under the ROC curve) \cite{Hanley1982}. Given the ranks of all objects in the testing set, the value of $AUC$ corresponds to the probability that a randomly chosen collected object is ranked higher than a randomly chosen un-collected object. After $N$ times independent comparisons of the resources received by a pair of collected and un-collected objects, the $AUC$ value is calculated by \cite{Zhou2009}
\begin{equation}
AUC = \frac{1}{m} \sum_{i=1}^{m} \frac{(N_1^{(i)} + 0.5N_2^{(i)})}{N^{(i)}},
\end{equation}
where $N_1^{(i)}$ denotes the times that the collected object has more resources than the uncollected object, and $N_2^{(i)}$ denotes the times that they have the same amount of resources for user $i$. Larger $AUC$ value suggests higher recommendation accuracy. The other one is the Ranking Score ($RS$) \cite{Zhou2007}. For a given user, $RS$ measures the relative ranking of a relevant object in the recommendation list. Formally, the value of $RS$ is calculated by \cite{Zhou2007}
\begin{equation}
RS = \frac{1}{|E^P|} \sum_{(i,\alpha) \in E^P} \frac{p_{\alpha}}{l_i},
\end{equation}
where $|E^P|$ is the size of the testing set, $p_\alpha$ is the position of object $\alpha$ in the ranking list of the recommendation, and $l_i$ is the number of uncollected objects of user $i$ in the training set. Smaller $RS$ value suggests higher accuracy of a recommendation algorithm.

We then introduce three $L$-dependent accuracy metrics: Precision, Recall, and $F_1$ \cite{Herlocker2004}. For all user, the average value of Precision $P(L)$ is calculated by
\begin{equation}
P(L) = \frac{1}{m}\sum_{i=1}^{m}\frac{d_i(L)}{L},
\end{equation}
where $d_i(L)$ is the number of recommended objects appeared in the testing set, and $L$ is the total number of recommended objects. The average value of Recall $R(L)$ is calculated by
\begin{equation}
R(L) = \frac{1}{m}\sum_{i=1}^{m}\frac{d_i(L)}{D(i)},
\end{equation}
where $D(i)$ is the total number of objects in the test set. Larger Precision and Recall suggest higher recommendation accuracy, however, the two measures are usually antagonistic since $P(L)$ usually decreases with $L$ while $R(L)$ usually grows with $L$. To balance both Precision and Recall, the $F_1$ metric is introduced. The average value of $F_1(L)$ is calculated by
\begin{equation}
F_1(L) = \frac{1}{m}\sum_{i=1}^{m} \frac{2P_i(L)}{P_i(L) + R_i(L)},
\end{equation}
where $P_i(L)$ is the value of Precision and $R_i(L)$ is the value of Recall for user $i$. Larger $F_1$ value suggests higher recommendation accuracy.

Diversity is an important metric in evaluating the variety of recommendations. One diversity metric is Hamming distance \cite{Zhou2008}. The average value of Hamming distance $H(L)$ is calculated by
\begin{equation}
H(L) = \frac{1}{m(m-1)}\sum_{i=1}^{m}\sum_{j=1}^{m}(1-\frac{C(i,j)}{L}),
\end{equation}
where $C(i,j)=| o_i^L \cap o_j^L |$ is the number of common objects in the lists of two users $i$ and $j$ with $L$ recommended objects. Larger $H(L)$ value suggests higher diversity. The other diversity metric is Intra-similarity \cite{ZhouT2009a}, which can be measured by the similarity between objects in the recommendation list. The average value of Intra-similarity $I(L)$ is calculated by
\begin{equation}
I(L) = \frac{1}{mL(L-1)}\sum_{i=1}^{m}\sum_{\substack{o_\alpha, o_\beta \in o_i^L \\ \alpha \neq \beta}}S_{\alpha\beta}^{Cos},
\label{eq.10}
\end{equation}
where $S_{\alpha \beta}^{Cos}$ is the cosine similarity between objects $\alpha$ and $\beta$ in the recommendation list $o_i^L$ of user $i$ with list's length $L$. Smaller $I(L)$ value suggests higher diversity of recommendations.

Novelty is a key metric for quantifying an algorithm's ability to generate novel (\emph{i.e.}, unpopular) and unexpected results \cite{Lu2012}. Here, we use the Popularity of the recommended objects to estimate the novelty of recommendations. The average Popularity $N(L)$ is calculated by
\begin{equation}
N(L) = \frac{1}{mL} \sum_{i=1}^m{\sum_{o_\alpha\in o_i^L}k_{\alpha}},
\end{equation}
where $k_\alpha$ is the degree of object $\alpha$ in the recommendation list. Smaller $N(L)$ value suggests higher novelty.

%%%%%%%%%%%%%%%%%%%%%%%%%%%%%%%%%%%%%%%%%%
\section{Results}
\label{sec4}

In this section, we will first present the results on analyzing the effects of the scaling parameter in the proposed CosRA+T method. Then, we will show the results on the recommendation performance of the CosRA+T method and the five benchmark methods. Next, we study how the length of recommendation list affects the algorithmic performance. Finally, we presents the results to help understand the mechanisms of these methods.

\subsection{The Impact of Scaling Parameter}

\begin{figure}[t]
  \centering
  \includegraphics[width=0.95\textwidth]{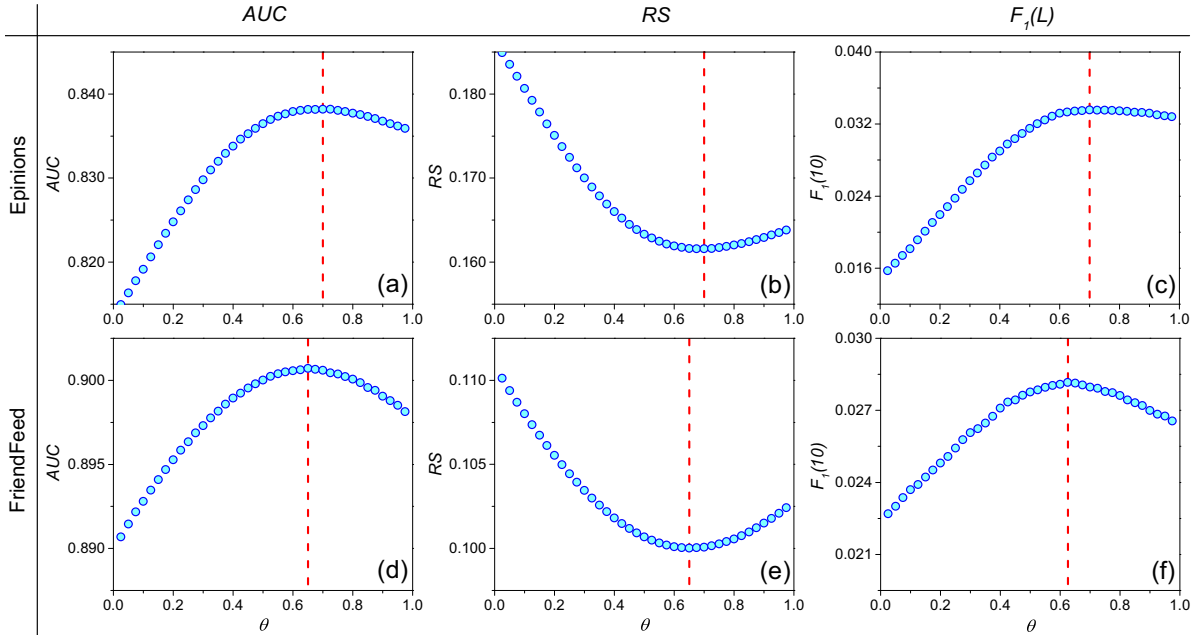}
  \caption{Results on the impact of the scaling parameter. (a-c) and (d-f) are for Epinions and FriendFeed, respectively. In the vertical-axis, three accuracy metrics are used, including $AUC$, $RS$, and $F_1(L)$ (from the left to the right column). In the horizontal-axis, the scaling parameter $\theta$ varies from 0 to 1. The vertical dashed red lines make the optimal value $\theta^*$, where the accuracy metrics reach their maximum. The length of the recommendation list is set as $L=10$ in calculating the $F_1(L)$ metric. The results are based on a 10-fold cross-validation and averaged over 20 independent realizations.}
  \label{Fig2}
\end{figure}

In the proposed CosRA+T method, a tunable scaling parameter $\theta$ is used to scale the resources that are received by the trusted users before the redistribution. The effects of users' trust relations on the resources-allocation processes increases as the scaling parameter $\theta$ decreases from 1 to 0. To explore the impact of the scaling parameter on the recommendation performance of the CosRA+T method, we vary $\theta$ from 0 to 1 and evaluate it on the Epinions and FriendFeed datasets by employing the three accuracy metrics: $AUC$, $RS$, and $F_1(L)$. Figure~\ref{Fig2} presents the corresponding results.

We observe that there seems to be an optimal value of the parameter $\theta$ for the best recommendation accuracy. This observation shows consistency on both Epinions and FriendFeed datasets under all the three accuracy metrics. Specifically, the values of $AUC$ and $F_1(L)$ gradually increase as the increase of $\theta$ while slightly decrease after $\theta$ crosses its optimal value $\theta^*$, which is around 0.70 for Epinions (see Figures~\ref{Fig2}(a) and \ref{Fig2}(c) for $AUC$ and $F_1(L)$, respectively) and around 0.65 for FriendFeed (see Figures~\ref{Fig2}(d) and \ref{Fig2}(f) for $AUC$ and $F_1(L)$, respectively). The value of $RS$ decreases (i.e., the recommendation accuracy increases) as $\theta$ increases before $\theta$ crossing its optimal value $\theta^* \approx 0.70$ for Epinions (see Figure~\ref{Fig2}(b)) and $\theta^* \approx 0.65$ for FriendFeed (see Figure~\ref{Fig2}(e)).

\begin{figure}[t]
  \centering
  \includegraphics[width=0.95\textwidth]{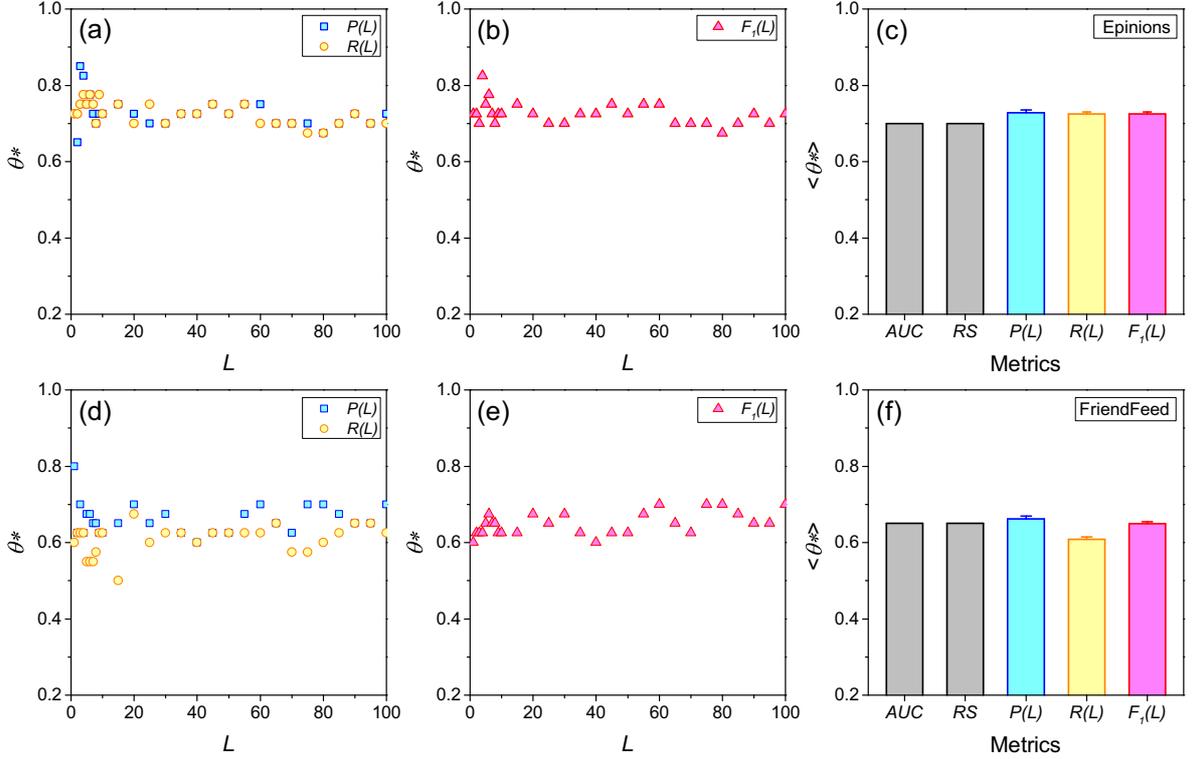}
  \caption{Results on the optimal value of the scaling parameter. (a-c) and (d-f) are for Epinions and FriendFeed, respectively. In (a) and (d), the optimal values $\theta^*$ in the vertical-axis are as a function of the recommendation list's length $L$ in the horizontal-axis for both $P(L)$ and $R(L)$. In (b) and (e), the results are for $F_1(L)$. In (c) and (f), the average optimal value $\left \langle \theta^* \right \rangle$ for each accuracy metric is presented with the error bar showing the standard error. The results are based on a 10-fold cross-validation and averaged over 20 independent realizations.}
  \label{Fig3}
\end{figure}

To investigate the universality of the optimal value $\theta^*$ and to better determine its critical value, we additionally employ the two $L$-dependent accuracy metrics, namely, Precision and Recall. By varying the recommendation list's length $L$, we explore how the optimal value $\theta^*$ changes with respect to $P(L)$, $R(L)$ and $F_1(L)$ on both datasets and present the results in Figure~\ref{Fig3}. We find that the optimal value $\theta^*$ of the scaling parameter are not sensitive to the recommendation list's length $L$. The optimal values $\theta^*$ are around 0.70 and 0.65 for Epinions and FriendFeed as indicated by the horizontal trends in Figures~\ref{Fig3}(a) and \ref{Fig3}(d), respectively. Similar trends are also observed for $F_1$ on both datasets (see Figure~\ref{Fig3}(b) for Epinions and Figure~\ref{Fig3}(e) for FriendFeed).

Further, we determine the optimal value $\theta^*$ by averaging the results under different $L$ for the three $L$-dependent accuracy metrics and present the results in Figures~\ref{Fig3}(a) and \ref{Fig3}(d) for Epinions and FriendFeed, respectively. Together, the results for the $L$-independent accuracy metrics ($AUC$ and $RS$) are also included. We found that the optimal values ($\left \langle \theta^* \right \rangle \approx 0.70 $ for Epinions and $\left \langle \theta^* \right \rangle \approx 0.65$ for FriendFeed) are very close to each other under several accuracy metrics on different datasets, suggesting the universality of the optimal scaling in CosRA+T for its best performance. The presence of the optimal scaling $\theta^*$ suggests that integrating the trust relations can enhance the recommendation performance, however, relying too much on it may result in the opposite. In other words, the integration of the trust relations is like the two sides of one coin, when it comes to the improvement of recommendation accuracy.

\subsection{Performance of Recommendation}

We further provide a more comprehensive evaluation of the proposed CosRA+T method and compare its performance with the five benchmark methods, namely, GR, UCF, HC, MD, and CosRA. We apply five accuracy evaluation metrics including the two $L$-independent metrics ($AUC$ and $RS$) and the three $L$-dependent metrics ($P(L)$, $R(L)$ and $F_1(L)$). We also employ two $L$-dependent diversity metrics ($H(L)$ and $I(L)$) and one popularity metric ($N(L)$). In the experiments, we set $L=10$ for all $L$-dependent metrics and analyze the impact of $L$ in the next section. In the comparisons with benchmark methods, CosRA+T method uses the optimal value $\theta^*$ for each dataset across all the evaluation metrics. Results of the algorithmic performance are summarized in Table~\ref{Tab2}.

As shown in the first five columns of Table~\ref{Tab2}, the proposed CosRA+T method outperforms all the five benchmark methods on both Epinions and FriendFeed datasets. Specifically, CosRA+T has remarkable advantages to GR, HC and UCF, as suggested by its larger $AUC$ value (0.8382 on Epinions and 0.9007 on FriendFeed). For the $L$-dependent metrics, CosRA+T has even better performance than these benchmark methods, for example, CosRA+T gives remarkably larger $P(L)$ and $R(L)$ values on both datasets. Also, CosRA+T gives competitive $F_1(L)$ value (0.0335) compared to that given by UCF (0.0259) and MD (0.0286) on Epinions. Moreover, the values of $RS$ given by CosRA+T are the smallest on both datasets, indicating that CosRA+T performs the best in recommendation accuracy among all the considered methods.

\begin{table}[t]
  \centering
  \footnotesize
  \caption{Results of the performance evaluation metrics after applying different recommendation methods on the Epinions and FriendFeed datasets.}
    \begin{tabular*}{\textwidth}{@{\extracolsep{\fill}}lcccccccc}
    \toprule
    Methods & $AUC$   & $RS$    & $P(L)$  & $R(L)$  & $F_1(L)$ & $H(L)$  & $I(L)$  & $N(L)$ \\
    \midrule
    Epinions &       &       &       &       &       &       &       &  \\
    GR    & 0.6974  & 0.3006  & 0.0094  & 0.0315  & 0.0144  & 0.1338  & 0.1389  & 308  \\
    HC    & 0.7845  & 0.2161  & 0.0052  & 0.0153  & 0.0077  & \textbf{0.9742} & \textbf{0.0245} & \textbf{5} \\
    MD    & 0.8256  & 0.1735  & 0.0189  & 0.0590  & 0.0286  & 0.6753  & 0.1140  & 235  \\
    UCF   & 0.8141  & 0.1844  & 0.0170  & 0.0537  & 0.0259  & 0.5748  & 0.1317  & 262  \\
    CosRA & 0.8356  & 0.1641  & 0.0221  & 0.0629  & 0.0327  & 0.9472  & 0.0900  & 107  \\
    CosRA+T & \textbf{0.8382} & \textbf{0.1616} & \textbf{0.0226} & \textbf{0.0651} & \textbf{0.0335} & 0.9544  & 0.0917  & 101  \\
    \midrule
    FriendFeed  &       &       &       &       &       &       &       &  \\
    GR    & 0.6058  & 0.3921  & 0.0050  & 0.0215  & 0.0081  & 0.0739  & 0.0935  & 172  \\
    HC    & 0.8833  & 0.1182  & 0.0088  & 0.0370  & 0.0142  & \textbf{0.9907} & \textbf{0.0542} & \textbf{11} \\
    MD    & 0.8925  & 0.1077  & 0.0163  & 0.0683  & 0.0263  & 0.9422  & 0.1195  & 73  \\
    UCF   & 0.8869  & 0.1133  & 0.0155  & 0.0661  & 0.0252  & 0.8857  & 0.1616  & 92  \\
    CosRA & 0.8978  & 0.1028  & 0.0167  & 0.0633  & 0.0265  & 0.9895  & 0.0890  & 35  \\
    CosRA+T & \textbf{0.9007} & \textbf{0.1000} & \textbf{0.0175} & \textbf{0.0693} & \textbf{0.0280} & 0.9899  & 0.1008  & 35  \\
    \bottomrule
    \end{tabular*}
    \begin{flushleft}
    \emph{Notes}: The length of the recommendation list is set as $L=10$. The scaling parameter $\theta$ in CosRA+T is set as its optimal value for each dataset. The results are based on a 10-fold cross-validation and averaged over 20 independent realizations. The best result of each metric is in bold.
    \end{flushleft}
  \label{Tab2}
\end{table}

The diversity metrics are shown in the sixth and seventh columns of Table~\ref{Tab2} for Hamming distance $H(L)$ and Intra-similarity $I(L)$, respectively. We notice that the recommendations given by HC have the best diversity as measured by diversity metrics on both Epinions and FriendFeed. Even though, CosRA+T still outperforms the other four benchmark methods as it gives larger $H(L)$ value (0.9544 on Epinions and 0.9899 on FriendFeed) and smaller $I(L)$ value (0.0917 on Epinions and 0.1008 on FriendFeed). The inferior of CosRA+T to HC is indeed very small, for example, the differences between $H(L)$ values are only 2.04\% and 0.08\% on Epinions and FriendFeed, respectively. Regarding the novelty of the recommendations, the performance of CosRA+T is also inspiring as it gives smaller $N(L)$ value than most benchmark methods except for HC (see the last column of Table~\ref{Tab2}). These results suggest that CosRA+T has overall larger accuracy, more diversity and better novelty.

\subsection{The Impact of Recommendation List's Length}

The length of recommendation list $L$ may affect the evaluation of recommendation performance under $L$-dependent metrics including three accuracy metrics ($P(L)$, $R(L)$ and $F_1(L)$), two diversity metrics ($H(L)$ and $I(L)$), and one novelty metric ($N(L)$). To this point, we explore how the length $L$ affects the performance of recommendation methods by varying $L$ from 1 to 100. The method of interest is CosRA+T, and benchmark methods are considered for comparison.

\begin{figure}[t]
  \centering
  \includegraphics[width=0.95\textwidth]{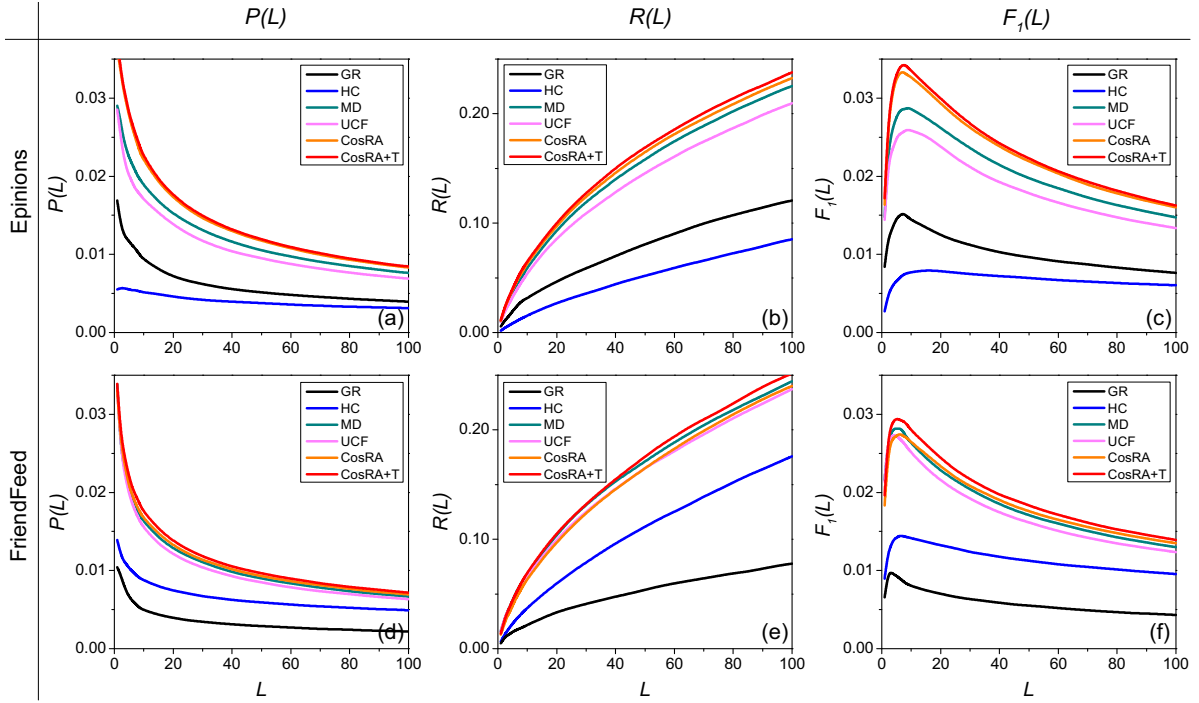}
  \caption{Results of the recommendation accuracy affected by the list's length. (a-c) and (d-f) are for Epinions and FriendFeed, respectively. In the vertical-axis, three accuracy metrics are respectively shown. In the horizontal-axis, the length of the recommendation list $L$ increases from 1 to 100. The scaling parameter $\theta$ in CosRA+T is set as its optimal value for each dataset. The results are based on a 10-fold cross-validation and averaged over 20 independent realizations.}
  \label{Fig4}
\end{figure}

Figure~\ref{Fig4} presents the results regarding the three accuracy metrics on both Epinions and FriendFeed datasets. We notice that, as the recommendation list's length $L$ increases, the values of $P(L)$ decreases (see Figures~\ref{Fig4}(a) and \ref{Fig4}(d)) while the values of $R(L)$ increases (Figures~\ref{Fig4}(b) and \ref{Fig4}(e)) for all the considered methods. The values of $F_1$ first increases and then decreases as $L$ increases, where the values of $F_1(L)$ reach their maximum at $L \approx 10$ on Epinions (see Figure~\ref{Fig4}(c)) and $L \approx 5$ on FriendFeed. Moreover, we find that the performance of CosRA+T is relatively better than other benchmark methods under different $L$, and its advantages are even remarkable when $L$ is around its optimal value, for example, $L^* \approx 10$ for Epinions. Further, we notice that CosRA+T has a relative improvement compared to the original CosRA, MD and UCF are competitive to each other, and HC and GR have the lowest accuracy.

\begin{figure}[t]
  \centering
  \includegraphics[width=0.95\textwidth]{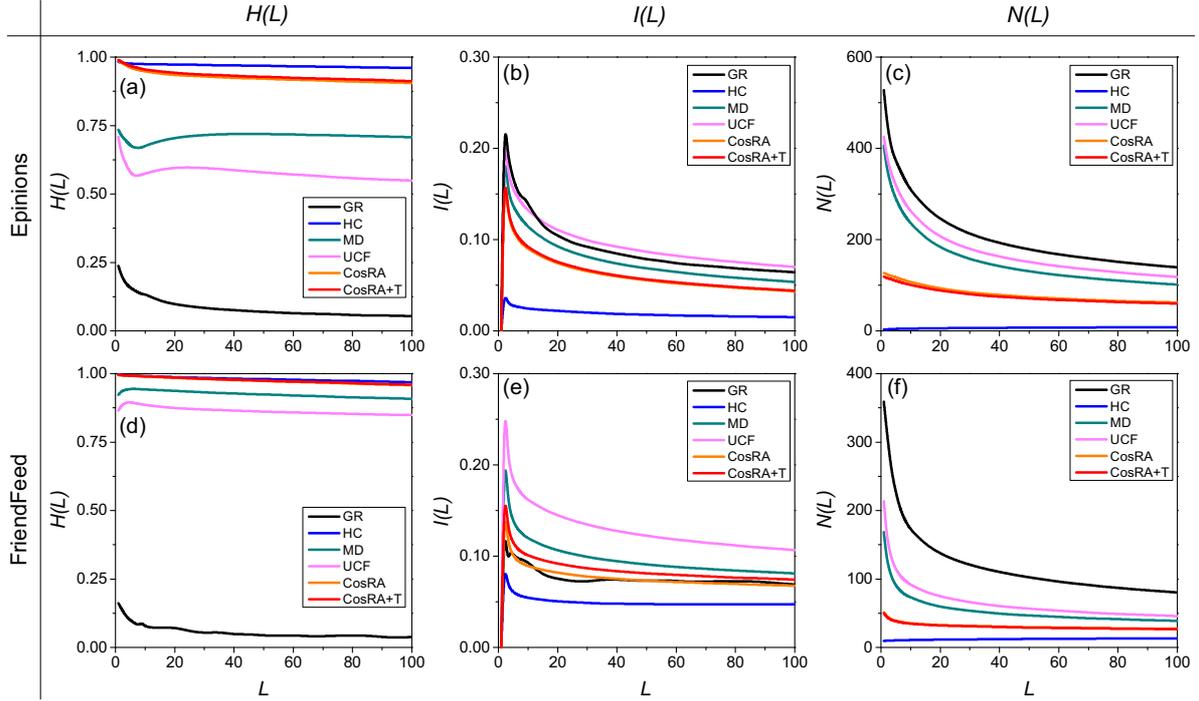}
  \caption{Results of the recommendation diversity and novelty affected by the list's length. (a-c) and (d-f) are for Epinions and FriendFeed, respectively. In the vertical-axis, two diversity metrics and one novelty metric are respectively shown. In the horizontal-axis, the length of the recommendation list $L$ increases from 1 to 100. The scaling parameter $\theta$ in CosRA+T is set as its optimal value for each dataset. The results are based on a 10-fold cross-validation and averaged over 20 independent realizations.}
  \label{Fig5}
\end{figure}

Figure~\ref{Fig5} presents the results regarding the two diversity metrics and the novelty metric. For the Hamming distance, the values of $H(L)$ slightly decrease as the increasing of $L$ on both Epinions and FriendFeed (see Figures~\ref{Fig5}(a) and \ref{Fig5}(d)). MD always gives the best results, followed close by CosRA and CosRA+T. UCF is remarkably inferior to MD, and GR performs the worst. For the Intra-similarity, the values of $I(L)$ first increase rapidly but then decreases slowly as $L$ increases (see Figures~\ref{Fig5}(b) and \ref{Fig5}(e)). UCF performs the worst as indicated by its largest $I(L)$ values. The performance of CosRA+T is ranked the second on Epinions but the middle on FriendFeed. HC outperforms all the other methods on both datasets as it gives the smallest $I(L)$ values. For the Novelty, the values of $N(L)$ decrease strongly as $L$ increases at the beginning but the decreases become slow thereafter (see Figures~\ref{Fig5}(c) and \ref{Fig5}(f)). We notice that CosRA+T outperforms most of the benchmark methods in the novelty metric except for HC, and the result is not sensitive to $L$. These results suggest that CosRA+T has relatively higher diversity and better novelty in the recommendations of uncollected objects.

\subsection{The Analysis of Mechanisms}

\begin{figure}[t]
  \centering
  \includegraphics[width=0.98\textwidth]{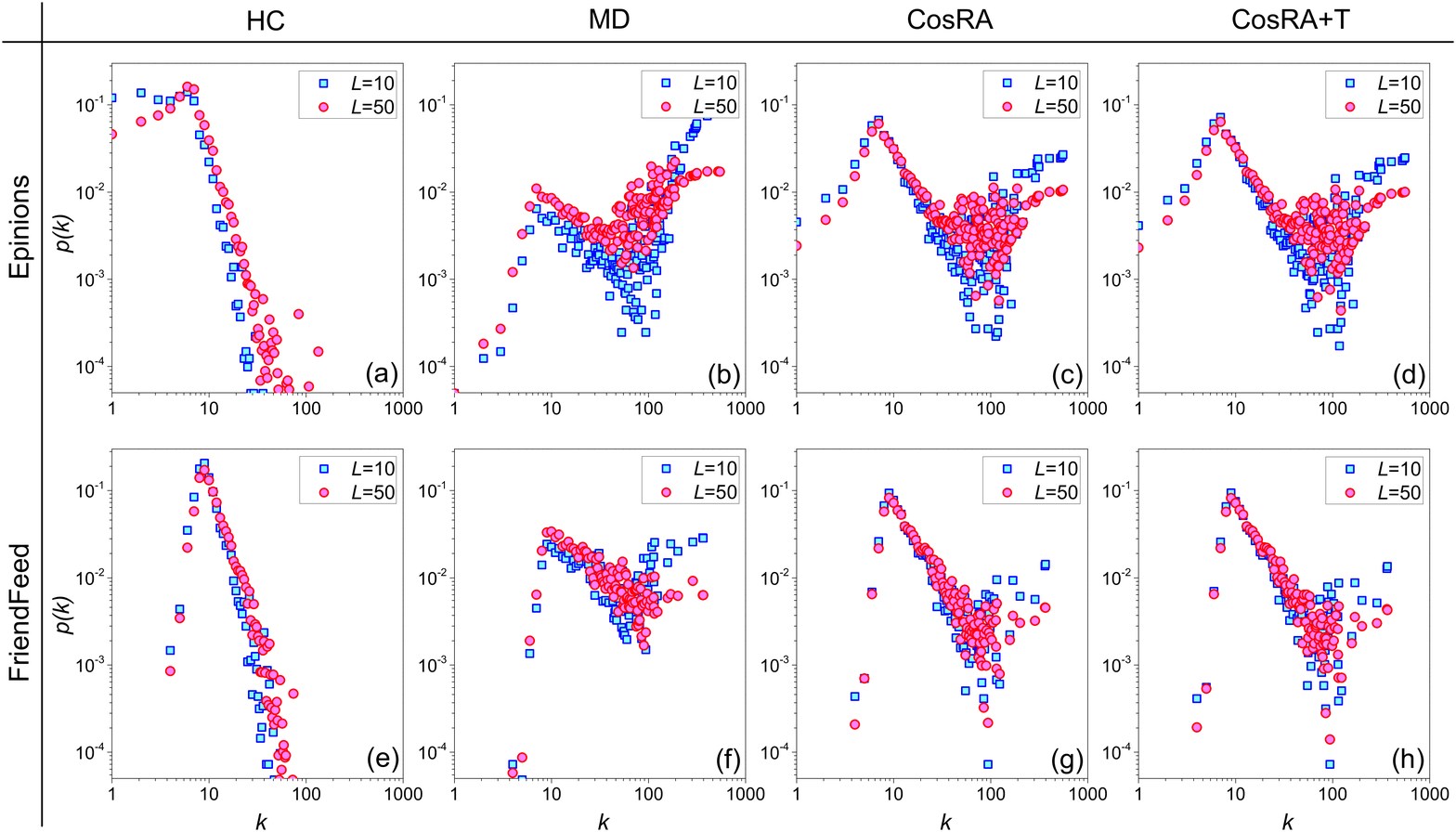}
  \caption{Degree distribution of the recommended objects after applying HC, MD, CosRA and CosRA+T methods on the Epinions and FriendFeed datasets. Results are shown for one realization on each dataset in log-log plot. Blue squares and red circles correspond to results under the recommendation list's length $L=10$ and $L=50$, respectively.}
  \label{Fig6}
\end{figure}

In order to better understand the mechanisms of the CosRA+T method, we further focus on the degree distributions of the recommended objects for all users. For the purpose of comparison, three benchmark methods (HC, MD and CosRA) are also considered in the study. Two different recommendation list's lengths ($L=10$ and $L=50$) are used. The scaling parameter $\theta$ in CosRA+T is set as its optimal value for each dataset. The results are presented in Figures~\ref{Fig6}(a-c) for Epinions and in Figures~\ref{Fig6}(d-f) for FriendFeed, respectively.

We notice that small-degree objects have a high probability to be recommended by HC (see Figures~\ref{Fig6}(a) and \ref{Fig6}(e) for Epinions and FriendFeed, respectively), as suggested by the relatively small degrees of the recommended objects. By comparison, there is a large change for large-degree objects being recommended by MD (see Figures~\ref{Fig6}(b) and \ref{Fig6}(f)) as we can observe that the degrees of its recommended objects are relatively large. These results suggest the strong trends and the potential bias of both HC and MD. When it comes to the recommendations, this issue may result in their remarkable disadvantages, for example, the low accuracy of HC and the poor novelty of MD.

By adopting a novel similarity index, CosRA fortunately finds a balance among the recommendation diversity and accuracy by recommending objects of large degree and small degree at the same time (see Figures~\ref{Fig6}(c) and \ref{Fig6}(g) for Epinions and FriendFeed, respectively). The main reasons are that, in the calculation of the CosRA index, the effects of popular objects with large degrees are restricted, and the effects of small-degree users are enhanced. Further, by introducing the scaling parameter to enlarge the resources received by trusted users before the redistribution processes, CosRA+T relatively improves the algorithmic performance by recommending both large-degree and small-degree objects (see Figures~\ref{Fig6}(d) and \ref{Fig6}(h)), especially when the recommendation list's length $L$ is around its optimal value.

%%%%%%%%%%%%%%%%%%%%%%%%%%%%%%%%%%%%%%%%%%
\section{Conclusion and Discussions}
\label{sec5}

In this paper, we explored the role of the trust relations among users in improving the performance of recommendation under the framework of network-based diffusion processes. Specifically, by introducing the trust relations into the original CosRA method, we proposed a trust-based recommendation method, named CosRA+T, in which the resources received by the trusted users are scaled by a tunable scaling parameter before being redistributed back to their collected objects in the networked resource redistribution process. We found an optimal scaling parameter for the proposed CosRA+T method, under which the method achieves the best accuracy in the recommendation. Interestingly, the optimal scaling values are very close to each other under different accuracy evaluation metrics on different rating and trust datasets. The result suggests the universality of the optimal scaling parameter in the proposed CosRA+T method for easy implementations.

Results of extensive experiments based on the two real-world rating and trust datasets, Epinions and FriendFeed, showed that the proposed CosRA+T method outperforms benchmark methods by giving a higher accuracy, an inspiring diversity and a well novelty in recommendations. Regarding the effects of the recommendation list's length on the performance evaluations under some parameter-dependent metrics (e.g., Precision and Recall), we found that the optimal lengths of the recommendation list are nearly the same on the same dataset for different methods while the optimal lengths may differ on different datasets. Finally, we tried to provide some insights to the mechanisms of some considered methods through presenting the degree distributions of their recommended objects. Results suggested that CosRA and CosRA+T balance well both small-degree and large-degree objects, leading to their better performance. Our work provides a promising step towards enhancing the recommendation performance by additionally considering users' social trust relations besides the traditionally used users' ratings.

The presented results are under some limitations on the datasets and the modeling process, which call for further improvements towards designing a better method that deals well with accuracy, diversity and novelty. The evaluation of algorithmic performance uses two rating datasets with trust relations, which are only represents of numerous real-world online rating platforms and socioeconomic systems \cite{Gao2016}. It would be an improvement if recommendation methods could be comprehensively tested and compared on various datasets and even on real platforms, focusing on how different recommendation algorithms affect the long-term evolution of online systems \cite{Zhao2013}. Moreover, the proposed method uses a simple way to enhance the resources received by trust users by a scaling parameter before the redistribution in the network-based diffusion processes. Yet, other possible realizations of introducing the trust relations into the rating-based methods under different frameworks could also be considered \cite{Zhou2007,Zhou2010}. Besides, not only social relationships (e.g., trust relations among users) but also users' features (e.g., online reputation of users \cite{Gao2015,Liu2017}) are critical to web-based recommender systems. As future works, a promising step is to combine social trust information and users' reputations \cite{Liao2014,Gao2017d} to further improve the performance of personalized recommendation algorithms for real-world applications.

%%%%%%%%%%%%%%%%%%%%%%%%%%%%%%%%%%%%%%%%%%
\section*{Acknowledgments}
The authors thank Shi-Min Cai, Qian-Ming Zhang and Tao Zhou for helpful discussions. This work was partially supported by the National Natural Science Foundation of China (Grant Nos. 61673086 and 61703074).

%%%%%%%%%%%%%%%%%%%%%%%%%%%%%%%%%%%%%%%%%%

%\bibliographystyle{elsarticle-num}
%\bibliography{ref}

%%%%%%%%%%%%%%%%%%%%%%%%%%%%%%%%%%%%%%%%%%
\end{document}